\documentclass[final,3p,times,twocolumn]{elsarticle}
\usepackage{natbib}
\usepackage{graphicx}

\newcommand{\dir}{Figs}

\newcommand{\veof}{v_{_{\mbox{\tiny EOF}}}}
\newcommand{\mueof}{\mu_{_{\mbox{\tiny EOF}}}}

\newcommand{\ack}{ {\bf Acknowledgment} }

\journal{Computer Physics Communication}

\begin{document}

\begin{frontmatter}
\title{Mesoscopic Simulations of Electroosmotic Flow and Electrophoresis in Nanochannels}
\author[A]{Jens Smiatek},
\author[B]{Friederike Schmid},
\address[A]{Institut f\"ur Physikalische Chemie, Universit\"at M\"unster, Germany}
\address[B]{Institut f\"ur Physik, Universit\"at Bielefeld, Germany}
\ead{e-mail: Friederike.Schmid@Uni-Mainz.DE}

\begin{abstract}
We review recent dissipative particle dynamics (DPD) simulations of electrolyte flow 
in nanochannels.
A method is presented by which the slip length $\delta_B$ at the 
channel boundaries can be tuned systematically from negative to infinity 
by introducing suitably adjusted wall-fluid friction forces. Using this
method, we study electroosmotic flow (EOF) in nanochannels for varying 
surface slip conditions and fluids of different ionic strength. Analytic expressions
for the flow profiles are derived from the Stokes equation, which are in 
good agreement with the numerical results. Finally, we investigate the influence 
of EOF on the effective mobility of polyelectrolytes in nanochannels. 
The relevant quantity characterizing the effect of slippage is found to 
be the dimensionless quantity $\kappa \delta_B$, where $1/\kappa$ is an 
effective electrostatic screening length at the channel boundaries. 
\end{abstract}
\end{frontmatter}

\section{Introduction}

Microfluidic devices like bio-MEMS (micro-electronical-mechanical-systems) and
bio-NEMS (nano-electronical-mechanical-systems) are attracting growing interest
due to their huge potential in bio- and nanotechnology, {\em e.g.}, for 
analyzing and manipulating tiny samples. Due to the large 
surface-to-volume ratios in nanoconfined systems, the flow in such devices is 
strongly influenced by the specific properties of the boundaries, {\em i.e.},
by surface characteristics like the wetting behavior and/or slippage.

If electric fields are involved, one particularly important mechanism is 
electroosmotic transport: In contact with a liquid, many materials commonly 
used in nanotechnology ({\em e.g.}, polydimethylsiloxane (PDMS)) acquire 
surface charges due to the ionization of surface groups \cite{Israelachvili}. 
Surfaces are thus often covered by a compensating counterion layer \cite{Hunter}. 
If one applies an external electric field, the ions are driven in one direction
and drag the surrounding fluid along, thus creating the so-called
electroosmotic flow (EOF). This electrokinetic effect has numerous consequences: 
It alters drastically the migration dynamics of mesoscopic objects like 
polyelectrolytes or colloids \cite{Viovy00}. In microchannels, the EOF 
generated at the channel walls induces a total net flow, which is technologically 
attractive because it can be controlled and manipulated more easily on the 
submicrometer scale than pressure- or shear-driven flow. 

One important application of microchannels is to use them for
separating different fragments of biological molecules like DNA by 
their length for sequencing or further manipulation.
In free solution, the electrophoretic mobility of high molecular weight 
polyelectrolytes is length independent \cite{Viovy00}. Electrophoretic
separation methods therefore often introduce the  samples into micro- or 
nanostructured environments, {\em e.g.}, disordered gels (in gel electrophoresis),
or structured microchannels \cite{Viovy00,Iki96,Roer97,effenhauser97,bader99,han00,
han02,huang02,duong03,ros04,Mathe07}. 
The migration behavior of the molecules in such
a setup results from a complex interplay of electrostatics, hydrodynamics, and
confinement effects on the molecules, and modeling all of this in full detail is
computationally challenging. Fortunately, it turns out that due to a combination of subtle 
screening effects -- which are not yet fully understood -- simple implicit-solvent 
Brownian Dynamics simulations that altogether neglect electrostatic and hydrodynamic 
interactions give results that are in qualitative and semiquantitative agreement with 
experiments \cite{Viovy00,Streek04,Streek05}. Nevertheless, such simplified 
treatments miss important and interesting physics, and simulations with explicit 
solvent and explicit charges are clearly desirable.

Here we present such explicit simulations of EOF and polyelectrolyte 
electrophoresis in nanochannels \cite{Smiatek09,Smiatek10},
using Dissipative Particle Dynamics (DPD) \cite{Hoo92,Esp95}, which is a popular
mesoscopic simulation method. Particular emphasis is put on the 
role of the channel boundaries, {\em i.e.}, on the effect of slippage and the 
electrical double layer at the surface. 

The remainder of the paper is organized as follows: After some theoretical
considerations, we describe the simulation model and method, with special focus 
on our method to implement variable hydrodynamic boundaries. Then we discuss the 
results, first for the simpler situation of a charged channel which only contains 
counterions, and finally for the full problem (electrolyte of 
varying ionic strength plus polyelectrolyte plus counterions). The results are 
compared with theory whereever possible. We conclude with a brief summary.

\section{Theoretical considerations: EOF in slit channels}
\label{sec:theory}

We consider a planar slit channel with identical walls at $z = \pm L/2$, exposed 
to an external electric field $E_x$ in the $x$ direction. The electrostatic potential 
$\Phi$ then takes the general form $\Phi(x,y,z) = \psi(z) + E_x \: x + \mbox{const}$, 
where we can set $\psi(0)=0$ for simplicity. Comparing the Poisson equation for the
electrostatic potential $\psi$ with the Stokes equation and exploiting the
symmetry of the channel, one finds \cite{Smiatek10}
\begin{equation}
 \label{eq:vx}
 v_x(z) = \psi(z) \cdot {\epsilon_r \: E_x}/{\eta_s}  + \veof,
\end{equation}
where $\epsilon_r$ is the dielectric constant, $\eta_s$ the shear viscosity of
the fluid, and $\veof = v_x(0)$ is an integration constant.

On the nanoscale, the appropriate hydrodynamic boundary condition at the
channel walls is the partial-slip boundary condition
\begin{equation}
\label{eq:partial_slip}
\delta_B \: \: \partial_{z} v(x)|_{{z}_B} =  v_{x}(z)|_{{z}_B},
\end{equation}
where $v_x(z)$ denotes the component of the velocity in x-direction evaluated at 
the position ${z}_B$ of an effective ``hydrodynamic boundary'' position, and 
the second effective parameter, the slip length $\delta_B$, characterizes
the amount of slippage at the surface. Inserting Eq. (\ref{eq:partial_slip})
into Eq.~(\ref{eq:vx}), we finally obtain the following simple
expression for the electroosmotic mobility,
\begin{equation}
 \label{eq:mueof}
 \mueof = {\veof}/{E_x}
 = \mueof^0 \: (1 + \kappa \: \delta_B),
\end{equation}
where we have defined an inverse 'surface screening length' \cite{Smiatek10}
\begin{equation}
 \label{eq:kappa}
 \kappa := \mp {\partial_z \psi}/{\psi} \large|_{z=\pm z_B},
\end{equation}
and $\mueof^0$ is the well-known Smoluchowski result \cite{Hunter} for the
electroosmotic mobility at sticky walls
$ \mueof^0 = - \epsilon_r \: \psi(z_B)/ \eta_s $.
A similar result can be derived within the linearized Debye-H\"uckel theory 
with $\kappa = \kappa_D$ (the Debye-H\"uckel length) \cite{Joly04}.

\section{Simulation Method}
\label{sec:method}

We study fluids that are driven through planar slit channels in the $x$-direction
by external fields, applying periodic boundaries in the $x$ and $y$ dimension 
and repulsive (charged) walls in the $z$ direction. The systems contained 
varying amounts of charged ions (anions and cations), plus possibly a
polyelectrolyte chain. They were thermalized by the momentum conserving DPD
thermostat \cite{Hoo92,Esp95}, where the forces acting on particles $i$ are given by
a sum
$
{\vec{F}}_{i}^{DPD}={\vec{F}}_i^C + \sum_{j\not={i}}({\vec{F}}_{ij}^{D}+{\vec{F}}_{ij}^{R})
$
of standard conservative contributions ${\vec{F}}_i^{C}$, dissipative forces 
$\vec{F}_{ij}^{D}$, and random forces $\vec{F}_{ij}^{R}$ with
\begin{eqnarray}
\label{eq:DPD}
  {\vec{F}}_{ij}^{D} &=&
  -\gamma_{DPD}\:\omega(r_{ij})\: (\hat{r}_{ij}\cdot{\vec{v}}_{ij}) \: \hat{r}_{ij} \\
\label{eq:random}
  {\vec{F}}_{ij}^{R} &=& \sqrt{2 \: \gamma_{DPD} \: k_B T \: \omega(r_{ij})} \: \:
 \check{\zeta}_{ij}\hat{r}_{ij}.
\end{eqnarray}
Here $\omega(r)$ is an arbitrary weight function with finite range $r_c$ (chosen linear in 
our case, $\omega(r) = 1-r/r_c$ for $r < r_c$), $\gamma_{DPD}$ is a friction coefficient,
$\hat{r}_{ij} = \vec{r}_{ij}/r_{ij}$ the unit vector in the direction of particle $j$,
$T$ the temperature, $k_B$ the Boltzmann factor, and 
$\check{\zeta}_{ij}=\check{\zeta}_{ji}$ are uncorrelated Gaussian 
distributed random variables with zero mean and unit variance. In DPD
simulations, the conservative forces are often taken to have a certain
soft shape. Here we only use the DPD {\em thermostat} as described above.
All simulations have been carried out with extensions of the freely available
software package {\sf{ESPResSo}} \cite{Espresso1}.

\subsection{Simulation Model}
\label{sec:model}

All particles, solvent, ions, and chain monomers, are modeled explicitly, and
have the same mass $m$ for simplicity.
Ions and monomers repel each other with a soft repulsive Weeks-Chandler-Anderson
(WCA) potential \cite{WCA} of range $\sigma$ and amplitude $\epsilon$. The same 
potential acts between particles and the walls. In addition, chain monomers are 
connected by harmonic springs
$ U_{harmonic}=\frac{1}{2}{k}(r_{ij}-r_0)^2 $
with the spring constant $k=25 \epsilon/\sigma^2$ and $r_0=1.0\sigma$.  Neutral
solvent particles have no conservative interactions except with the walls.
The wall contains immobilized, negatively charged particles at random
positions. Every second monomer on the polyelectrolyte carries a negative charge.
All charges are monovalent, and the system as a whole is electroneutral. 
Charged particles interact with each other {\em via} a Coulomb potential with the 
Bjerrum length $\lambda_B=e^2/4\pi\epsilon_r k_BT=1.0\sigma$, and also with an
external electric field $E_x=-1.0 \epsilon/e\sigma$. 
Specifically, we show here results for systems with channel width $8 \sigma$ and
a surface charge density $\sigma_A=-0.208e\sigma^{-2}$, which is found to correspond
to the 'weak-coupling regime' \cite{Smiatek09}, {\em i.e.}, the regime where 
the Poisson-Boltzmann theory is valid. The total counterion density was roughly 
$\rho_{\mbox{\tiny counter}} = 0.06\sigma^{-3}$ and the salt density varied between
$\rho_s=$0.05625, 0.0375, 0.03, 0.025, and $0.015 \sigma^{-3}$. In molar units, this
corresponds to 0.272, 0.181, 0.145, 0.121 and 0.072 mol/l, if we identify
$\lambda_B \approx 0.7$ nm, {\em i.e.}, the Bjerrum length in water at room temperature
\cite{Viovy00}.

\subsection{Tunable Slip boundaries}
\label{sec:slip}

\begin{figure}[t]
\vspace{\baselineskip}
\centerline{
\includegraphics[width=0.42\textwidth]{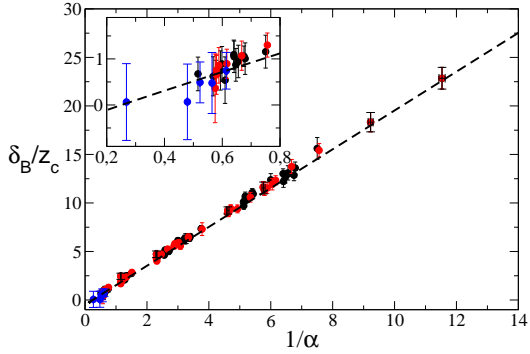}
}
\vspace{-0.5\baselineskip}
 \caption{
\label{fig:sliplength}
Slip length $\delta_B$ in units of $z_c$ vs.\ $\alpha$ for varying values of the
   parameter triplet $(\rho,\gamma_{DPD},\gamma_L)$ (in units of $\sigma^{-3}$ or
   $\sqrt{m \epsilon}/\sigma$, respectively).
    Black: series with $\rho$ fixed: (3.75 , 2-10, 0.1-1).
    Red: series with $\gamma_{DPD}$ fixed: (3.75-12.5, 2, 0.1-1).
    Blue: selected triplet values: (6.35,5,1),(5,5,1), (11.25,2,1.1), (11.25,2,1.2), (3.75,10,2.5).
    Dashed black line: Theory (Eq.~(\protect\ref{eq:db_linear}))
    The inset shows a blowup of the same data.
    After Ref. \protect\cite{Smiatek08}.
}
\end{figure}

To realize arbitrary hydrodynamic boundary condition at the walls, we introduce 
an additional coordinate-dependent viscous force that mimicks a 
wall/fluid friction \cite{Smiatek08}
\begin{equation}
\label{eq:langevin}
  {\vec{F}}_i^{L}={\vec{F}}_i^{D}+{\vec{F}}_i^{R}
\end{equation}
with a dissipative contribution
\begin{equation}
  {\vec{F}}_i^{D}=-\gamma_L\: \omega_L(z) \: \: ({\vec{v}}_i-{\vec{v}}_{wall})
\end{equation}
coupling to the relative velocity $({\vec{v}}_i-{\vec{v}}_{wall})$ of the particle
with respect to the wall, and a stochastic force
\begin{equation}
  F_{i,\beta}^R= \sqrt{2 \gamma_L \: k_B T \: \omega_L(z)}\;\chi_{i,\beta}.
\end{equation}
Here $\beta$ runs over $\beta = x,y,z$, $\chi_{i,\beta}$ is a Gaussian distributed
random variable with mean zero and unit variance, and the weight function 
$\omega_L(z)$ is chosen $\omega_L(z) = 1 - z/z_c$ up to a cut-off distance $z_c$.  
The prefactor $\gamma_L$ sets the strength of the friction force and hence determines 
the value of the slip length.  With this approach it is possible to tune the slip
length $\delta_B$ systematically from full-slip to no-slip and even (small)
negative slip. Furthermore, one can use the Stokes equation to derive an 
analytical expression for the slip length $\delta_B$ as a function of the model 
parameters, giving \cite{Smiatek08}
\begin{equation}
\label{eq:db_linear} \frac{\delta_B}{z_c} =
  -1+\;\frac{1}{(3\alpha)^{1/3}}\;
  \frac{\Gamma\left(\frac{1}{3}\right)}{\Gamma\left(\frac{2}{3}\right)} \;
  \frac{I_{-2/3}\left(\frac{2\sqrt{\alpha}}{3}\right)}
  {I_{2/3}\left(\frac{2\sqrt{\alpha}}{3}\right)}
\end{equation}
where $\Gamma$ is the Gamma-Function, $I$ the modified Bessel function of
the first kind, and the dimensionless parameter 
$\alpha = z_c^2 \gamma_L \rho /\eta_s$ depends on the density $\rho$ and the 
shear viscosity $\eta_s$ of the fluid. To test this expression, we have studied
driven neutral fluids in our slit channels for the range of parameters 
$\gamma_L = (0.1$ -$ 5) \:\sqrt{m \epsilon}/\sigma$, 
$\rho = (3.75$-$12.5) \:\sigma^{-3}$, and 
$\gamma_{DPD} = (1$-$10) \sqrt{m \epsilon}/\sigma$. 
By performing Plane Poiseuille and Plane Couette flow simulations, one can
determine $\eta_s$, $\delta_B$, and the position of the hydrodynamic 
boundary independently \cite{Smiatek08}. As expected, the hydrodynamic
boundary is always close to the physical boundary, 
whereas the slip length varies over a wide range. Fig. \ref{fig:sliplength} 
shows that the results agree very nicely with the theoretical prediction, 
Eq. (\ref{eq:db_linear}) \cite{Smiatek08}.

\section{Results}
\label{sec:results}

\subsection{Counterion-induced electroosmotic flow}
\label{sec:CI-EOF}

\begin{figure}[t]
\vspace{\baselineskip}
\centerline{
\includegraphics[width=0.42\textwidth]{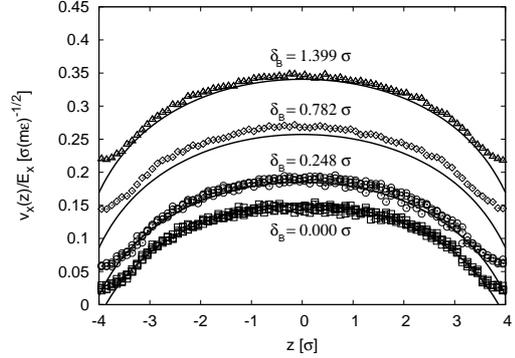}
}
\vspace{-0.5\baselineskip}
 \caption{
\label{fig:counterion_EOF}
Flow profiles for counterion-induced flow at field strengths 
$E_x=0.8-1.0k_B T/e\sigma$ for varying slip lengths in the 
weak coupling regime. The straight lines represent the
theoretical prediction of Eqn.(\ref{eq:ana_EOF}).
}
\end{figure}

We begin with discussing the ''simpler'' situation without polyelectrolyte
and salt, where the fluid only contains the counterions of the charges in 
the wall. In the regime of validity of the Poisson-Boltzmann equation
(the weak coupling limit) the potential distribution $\psi(z)$ can be calculated 
analytically, giving \cite{Israelachvili} $\psi(z) \propto \log(\cos^2(\kappa_c z))$,
which results in the counterion distribution $\rho_c(z) = \rho_0 /\cos^2(\kappa_c z)$
with the screening constant $\kappa_c^2 = e^2\rho_0/2\epsilon_r k_B T$.
The parameter $\rho_0$ (the counterion density in the middle of the channel)
is set by the electroneutrality requirement, {\em i.e.}, the integrated 
counterion density $\int {\rm d}z \: \rho_c(z)$ must equal the surface charge density 
$\sigma_A$. In our case, the corresponding calculation gives $\rho_0 = 0.0174 \sigma^{-3}$, 
in accordance with the numerical value, 
$\rho_0 = (0.0176 \pm 0.0001) \sigma^{-3}$ \cite{Smiatek09}.
By virtue of Eq.~(\ref{eq:vx}) combined with (\ref{eq:partial_slip}),
one finally obtains the explicit expression
\begin{eqnarray}
\nonumber
  v_x(z) &=&
  \frac{e }{4 \pi \lambda_B Z \eta_s}E_x \: \Big( \: 
    \log \big( \frac{\cos^2(\kappa_c z_B)}{\cos^2(\kappa_c z)} \big)  \\&&
     + 2 \: \kappa_c \delta_B\tan(\kappa_c z_B) \: \Big)
  \label{eq:ana_EOF}
\end{eqnarray}
for the flow profile, where we have expressed $\epsilon_r$ in terms
of the Bjerrum length $\lambda_B$. Fig.~\ref{fig:counterion_EOF} compares
our numerical results for varying surface characteristics (slip lengths)
and field amplitudes with the theoretical prediction, Eq. (\ref{eq:ana_EOF}). 
The theory describes the data very nicely, without any fit parameter.

\subsection{The full problem}

We are now ready to consider the full problem, {\em i.e.}, a system containing 
polyelectrolyte, counterions, and varying amounts of salt ions, 
$\rho_s = 0.015-0.056 \sigma^{-3}$.
\begin{figure}[h]
\vspace{\baselineskip}
\centerline{
\includegraphics[width=0.42\textwidth]{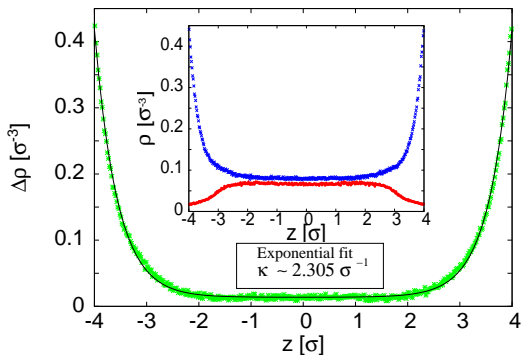}
}
\vspace{-0.5\baselineskip}
 \caption{
\label{fig:ions_all}
  {\bf Inset:} Distribution $\rho_c$ of cations (salt cations and counterions, 
  blue symbols) and anions $\rho_a$ (salt anions, red symbols) for
  a system containing polyelectrolyte at the salt concentration
  $\rho_s = 0.05625 \sigma^{-3}$.
 {\bf Main frame:} Corresponding ionic difference profile $\Delta \rho=\rho_c-\rho_a$ 
  The exponential fit (black line), gives the effective inverse screening length 
  $\kappa=2.305\pm0.025\sigma^{-1}$.
}
\end{figure}

Fig.~\ref{fig:ions_all} shows the ion distribution profiles for one salt concentration. 
The profiles of the ionic difference exhibit an
exponential behavior $f(z) = A(e^{-\kappa z} + e^{\kappa z} + c)$,
although the fitted screening parameter $\kappa=(2.305 \pm 0.025) \sigma^{-1}$
does not agree well with the Debye-H\"uckel screening length, 
$\kappa_D = 1.21 \sigma^{-1}$. Similar observations were made at all other
salt concentrations: For the surface charge $\sigma_s = 0.208 \sigma^{-2}$,
the linearized Debye-H\"uckel theory is not valid. Nevertheless, a
well-defined surface screening length $\kappa$ can be extracted from the data
by a simple exponential fit.

\begin{figure}[t]
\vspace{\baselineskip}
\centerline{
\includegraphics[width=0.42\textwidth]{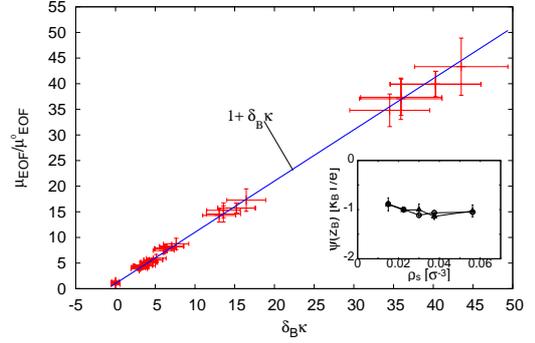}
}
\vspace{-0.5\baselineskip}
 \caption{
\label{fig:EOF_all}
Ratio $\mu_{_{EOF}}/ \mu_{_{0,EOF}}$ plotted against $\delta_B\kappa$
  for salt concentrations $\rho_s$ ranging in $0.015--0.056 \sigma^{-3}$
  and various slip lengths. The blue line is the theoretical prediction of 
  Eq.~(\ref{eq:mueof}) with slope $1+\delta_B\kappa$.
  {\bf Inset:} Surface potential $\psi(z_B)$ as a function of salt concentration
  $\rho_s$.
}
\end{figure}

Next we discuss the electroosmotic flow in these channels. Fig.~\ref{fig:EOF_all} 
compiles our numerical results for the EOF mobility for all salt concentrations 
and slip lengths. They are in very good agreement with the theoretical prediction 
of Eq.~(\ref{eq:mueof}), where $\mu_{_{EOF}}^0$ has been determined independently 
by a linear regression for each salt concentration. It is worth noting that
the presence of the polyelectrolyte does not perturb the amplitude
of the electroosmotic flow. 

Finally, we consider the effective migration of the polyelectrolyte in
the electric field. It results from a combination of two effects: the 'bare' 
electrophoresis relative to the surrounding fluid, and the convective transport 
by the electroosmotic flow. In many situations of interest, one can argue that 
these two contributions simply add up \cite{Streek04,Smiatek10}. 
The total mobility $\mu_t$ is then given by $\mu_t = \mueof +\mu_e$,
where $\mu_e$ is the electrophoretic mobility of the polyelectrolyte in
a hypothetical fluid at rest, and by virtue of Eq.~(\ref{eq:mueof}), it
can be expressed in terms of the electroosmotic mobility $\mueof$ as
\begin{equation}
\label{eq:eof_mobp}
  \frac{\mu_t}{\mueof}=1+\frac{\mu_e}{\mueof^0(1+ \kappa \: \delta_B)},
\end{equation}
where the ratio $\mu_e/\mueof^0$ is expected to depend only weakly on the ionic 
strength of the electrolyte and the slip length of the surface. The main effect 
of slippage is incorporated in the factor $(1 + \kappa \: \delta_B)^{-1}$ 
\cite{Smiatek10}.

\begin{figure}[t]
\vspace{\baselineskip}
\centerline{
\includegraphics[width=0.42\textwidth]{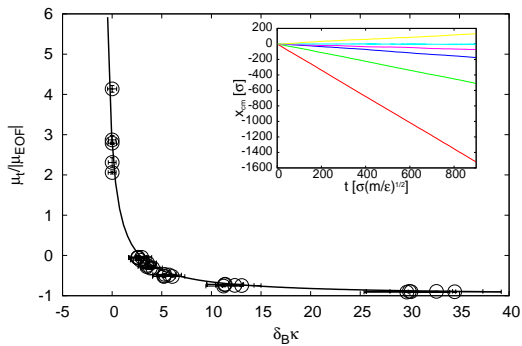}
}
\vspace{-0.5\baselineskip}
 \caption{
\label{fig:mobility}
Ratio $\mu_t/|\mu_{_{EOF}}|$ plotted against $\delta_B\kappa$ for all
  salt concentrations. The black line is the theoretical prediction of
  Eq.~(\ref{eq:eof_mobp}), with one single fit parameter
  $\mu_e/\mu_{_{EOF}}^0=-3.778\pm 0.128$. Negative values
  of $\mu_t/|\mu_{_{EOF}}|$ indicate absolute negative total mobilities 
  of the polyelectrolyte.
  {\bf Inset:}
  Total displacement of the polyelectrolytes center of mass for different
  boundary conditions at the salt concentration 
  $\rho_s=0.05625\sigma^{-3}$. The lines correspond from
  top to bottom
  to the slip lengths $\delta_B\approx (0.00, 1.292, 1.765, 2.626, 5.664,
  14.98)\sigma$. 
}
\end{figure}

Our numerical results for the total mobility of the polyelectrolyte for varying 
boundary conditions are presented in Fig.~\ref{fig:mobility}. They are in
excellent agreement with the theoretical prediction, Eq.~(\ref{eq:eof_mobp}),
with one single fitted ratio $\mu_e/\mu_{_{EOF}}^0=-3.778\pm 0.128$. For stick
boundaries ($\delta_B\approx 0$) one obtains ordinary behaviour where the 
polyelectrolyte follows the electric force acting on the monomers. In the presence
of slip, the absolute mobility may become negative if the  electroosmotic flow 
exceeds a critical value. This is because the immobile wall charges and the 
charges on the polyelectrolyte have the same sign, hence the directions of
the EOF and the bare electrophoresis are opposite. If the wall charges
and the polyelectrolyte charges are opposite, slippage effects should enhance 
the total mobility of the polyelectrolyte.

\section{Conclusions and Outlook}
\label{sec:conclusions}

We have presented mesoscopic DPD simulations of EOF and polyelectrolyte
electrophoresis in narrow microchannels, taking full account of hydrodynamic
and electrostatic interactions. Slippage effects massively
influence the electroosmotic flow and therefore the total mobility
of the polyelectrolyte. Under certain conditions, even a negative mobility
can be achieved, in agreement with recent experiments \cite{Mathe07}.  
All our numerical results are in good agreement with analytical expressions,
which were derived based on the Stokes equation.

In sum, our mesoscopic simulations indicate that the migration of polyelectrolytes
in nanochannels results from the interplay of electroosmotic, electrophoretic,
electrostatic and slippage effects. To describe the mobility adequately, all
of these factors need to be accounted for. From a technological point, 
the characteristics of the channel walls could be used to significantly enhance 
flow amplitudes, which offers the possibility to reduce the time needed for 
polymer migration or separation techniques. This could be an important aspect 
for future applications in microchannels or micropumps.

\ack We have benefitted from interactions and discussions with
Michael P. Allen, Christian Holm, Burkhard D\"unweg, Ulf D. Schiller, 
Marcello Sega, and Kai Grass. The simulations were carried out at
the Arminius Cluster PC$^2$ at Paderborn University,
the HLRS in Stuttgart and the NIC computing center in J\"ulich
for computer time. This work was funded by the Volkswagenstiftung.


\begin{thebibliography}{99}
\bibitem{Israelachvili} Israelachvili, J.: Intermolecular and Surface Forces. Academic Press,
  London (1991).
\bibitem{Hunter}
  Hunter, R.~J.: Foundations of Colloid Science, Vol.1. Clarendon Press, Oxford (1991).
\bibitem{Viovy00} Viovy, J.-L.: Rev.~Mod.~Phys. {\textbf{72}}, 813 (2000).
\bibitem{Iki96} Iki, N.; Kim, Y.; Yeung, E.~S.: Anal.~Chem. {\bf 68}, 4321 (1996).
\bibitem{Roer97}
  Roeraade, M.; Stjernstr{\"o}m, M. International Patent WO/1997/26531, {\bf 1997},
  avaible at {\em http://www.wipo.int}.
\bibitem{effenhauser97} Effenhauser, C.~S.; Bruin, G.~J.~M.; Paulus, A.:
  Electrophoresis {\bf 18}, 2203 (1997). 
\bibitem{bader99} Bader, J.~S. {\em et al}: PNAS {\bf 96}, 13165 (1999).
\bibitem{han00} Han, J.; Craighead, G.: Science {\bf 288}, 1026 (2000).
\bibitem{han02} Han, J.; Turner, S. W.; Craighead, G.: Phys.~Rev.~Lett. {\bf 83}, 1688 (2002).
\bibitem{huang02}
 Huang, L.~R. {\em et al}: Nature Biotechnology {\bf 20}, 1048 (2002).
\bibitem{duong03} Duong, T.~T. {\em et al}: Microelectronic Engineering {\bf 67}, 905 (2003).
\bibitem{ros04} Ros, A. {\em et al}: J. Biotechnology {\bf 112}, 65 (2004).
\bibitem{Mathe07} Mathe, J., Di Meglio, J.~-M., Tinland, B.: J.~Colloid Interface Sci., 
  \textbf{316}, 831 (2007).
\bibitem{Streek04} Streek, M. {\em et al}: J. Biotechnology {\bf 112}, 79 (2004).
\bibitem{Streek05} Streek, M. {\em et al}: Phys. Rev. E {\bf 71}, 11905 (2005).
\bibitem{Smiatek09} Smiatek, J. {\em et al}: J. Chem. Phys. {\bf 24}, 244702 (2009).
\bibitem{Smiatek10} Smiatek, J., Schmid, F.: J. Phys. Chem. B. {\textbf 114}, 6266 (2010)
\bibitem{Hoo92} Hoogerbrugge, P.~J., Koelman, J.~M.~V.~A.: Europhys.~Lett. {\textbf{19}},
  155, (1992).
\bibitem{Esp95}
  Espa\~nol, P., Warren, P.~B.: Europhys.~Lett. {\bf 30}, 191 (1995)
\bibitem{Joly04} Joly, L. {\em et al}: Phys. Rev. Lett. {\bf 93}, 257805 (2004).
\bibitem{Espresso1} Arnold, A. {\em et al}, Comp.~Phys.~Comm., {\bf{174}}, 704 (2005).
\bibitem{WCA} Weeks, J.~D.; Chandler, D.; Andersen, H.~C.: J.~Chem.~Phys. {\bf 54},
 5237 (1971).
\bibitem{Smiatek08}
  Smiatek, J., Allen, M.~P., Schmid, F.: Eur.~Phys.~J.~E, {\bf 26}, 115 (2008).
\end{thebibliography}
\end{document}